**Single trial ERP amplitudes reveal the time course of acquiring representations of novel faces in individual participants**

Werner Sommer

Humboldt-University at Berlin, Department of Psychology, Germany

Katarzyna Stapor

Silesian University of Technology, Institute of Computer Science, Gliwice, Poland

Grzegorz Kończak

University of Economics in Katowice, Department of Statistics, Econometrics and Mathematics, Poland

Krzysztof Kotowski

Silesian University of Technology, Department of Applied Informatics, Gliwice, Poland

Piotr Fabian

Silesian University of Technology, Department of Algorithmics and Software, Gliwice, Poland

Jeremi Ochab

Jagiellonian University, Department of Theory of Complex Systems, Krakow, Poland

Anna Bereś

Jagiellonian University, Department of Cognitive Neuroscience and Neuroergonomics , Krakow, Poland

Grażyna Ślusarczyk

Jagiellonian University, Department of Design and Computer Graphics, Krakow, Poland

Correspondence should be addressed to

Werner Sommer, Institut für Psychologie, Humboldt-Universität zu Berlin, Rudower Chaussee 18, 12489 Berlin, Germany

or

Katarzyna Stapor, Faculty of Automatic Control, Electronics and Computer Science, Department of Applied Informatics, Silesian University of Technology, Akademicka 16, 44-100 Gliwice, Poland

**Abstract**

The neural correlates of face individuation – the acquisition of memory representations for novel faces – have been studied only in coarse detail and disregarding individual differences between learners. In their seminal study, (Tanaka, Curran, Porterfield, & Collins, 2006) required the identification of a particular novel face across 70 trials and found that the N250 component in the ERP became more negative from the first to the second half of the experiment, where it reached a similar amplitude as a well-known face. We were unable to directly replicate this finding in our study when we used the original spilt of trials. However, when we applied a different split of trials we observed very similar changes in N250 amplitude. Then, we developed and applied a new two-step explorative-confirmative non-parametric method based on permutation testing to determine the time course of face individuation in individual participants based on single trial N250 amplitudes. We show that the assumption of a steep initial increase of N250 amplitude across multiple presentations of the target face, followed by a plateau, yields plausible results in fitting linear trends for most participants. The transition point from initial acquisition to the plateau phase differed strongly between participants and tended to be earlier when performance in target face recognition was better. Hence, face individuation may be accounted for by a biphasic process of early, fast acquisition, followed by a slower, asymptotic consolidation or maintenance phase. The current approach might be fruitfully applied to further investigations into face individuation and their neural correlates.

**KEYWORDS**: Face learning, ERP, N250, single-trial analysis, modeling

# Single trial ERP amplitudes reveal the time course of acquiring

## representations of novel faces in individual participants

## 1   INTRODUCTION

The recognition of the faces of individual persons is crucial for social life. Although it is often very easy to recognize familiar faces, it is much harder for unfamiliar faces (Young & Burton, 2017) especially from single instances (Logie, Baddeley & Woodhead, 1987). Therefore, the development of stable traces of faces over repeated encounters despite image variability is being seen as outstanding research question (Andrews, Burton, Schweinberger, & Wiese, 2017). In the present study we suggest a new method to trace the emergence of a neural correlate of face familiarity trial by trial and to characterize this development at the level of individual learners.

ERPs, derived from the EEG, provide a number of components that are sensitive to the cognitive processing of faces. Following the P1 component, which is mainly sensitive to domain-general visual processes, faces elicit a prominent N1 component peaking around 170 ms post stimulus onset (N170; Eimer, 2011). The N170 is larger to faces than to most other objects and increases in amplitude and latency when the structure of a face is hard to perceive; however, it has rarely been found to be sensitive to individual faces or to face familiarity (Zimmermann & Eimer, 2013; Tanaka et al., 2006; Kaufmann, Schweinberger, & Burton 2009).

In contrast to the N170, the subsequent N250 component has been shown to be familiarity-sensitive. The N250 is negative at occipito-central sites and positive at frontal sites, and is thought to be generated in or near the fusiform gyrus (Kaufmann et al., 2009; Schweinberger, Kaufmann, Moratti, Keil, & Burton, 2007). The N250 increases if a face image is the same as an immediately preceding face as compared to when it is different (Schweinberger, Pfütze, & Sommer, 1995). Importantly, this early repetition or priming effect – also termed

N250r – is more pronounced for familiar faces than for unfamiliar faces. With respect to face individuation, the N250r has been shown to increase when a face grows more familiar (Zimmermann & Eimer, 2013). Other studies showed that also the N250 component itself increases when unfamiliar faces are presented repeatedly Tanaka et al. (2006) and that similar effects hold when experts view non-face objects in their domain of expertise (Pierce, Scott, Boddington, Droucker, Curran, & Tanaka, 2011). Supporting the findings of Tanaka et al. (2006), Schulz, Kaufmann, Kurt, and Schweinberger (2012) showed larger N250 amplitudes to explicitly learned than to novel faces.  Andrews et al. (2017) reported that the increase of N250 holds also when during the ERP test different images were used than during (incidental) learning. Hence it has been concluded that the N250 is associated with the processing of familiar objects at the subordinate representational level (Scott, Tanaka, Sheinberg, & Curran, 2008; Tanaka et al., 2006; Wiese, Tüttenberg, Ingram, Chan, Gurbuz, Burton, & Young, 2019)).

Previous studies, demonstrating the relationship of the N250 with face (or object) familiarity have either compared familiar with (different) unfamiliar faces or familiar and unfamiliar objects (e.g., cars, dogs; Pierce et al., 2011) or they have compared the N250 across two or three consecutive blocks of trials (Tacikowski, Jednoróg, Marchewka, & Nowicka 2011; Tanaka et al., 2006; Zimmermann & Eimer, 2013) as stimuli increased in familiarity. Kaufmann et al. (2009) presented (several) faces in four blocks which had been seen in videos before and found an increase of N250 after the first block. While the averaging of several responses within a block of presentations and across participants offers the advantage of improved signal to noise ratio, it does not allow to study the time course of the learning process within such blocks. Furthermore previous studies disregard individual differences which are prominent in face cognition (e.g., Wilhelm, Herzmann, Kunina, Danthiir, Schacht, & Sommer, 2010). In the present study we suggest a procedure that will allow more fine-grained analyses.

Starting point of the present study was the Joe/No Joe task of (Tanaka et al., 2006), where eleven unfamiliar faces and the participant's own face were presented multiple times. One of the unfamiliar faces was designated as "Joe's" face (target face). When the responses to "Joe" presentations were averaged separately for the first half and the second half of the experiment (35 trials each), the N250 amplitude within a ROI of 12 electrodes at posterior areas of the scalp in the time range 230-320 ms post-stimulus was larger (more negative) in the second than in the first half of the experiment and became indistinguishable from the N250 to the highly familiar picture of the participant's own face.

In  the first step of the present study, we attempted to replicate the results of (Tanaka et al., 2006), repeating their block-wise ERP averaging as a starting point. In the second step we explored in detail the process of face learning for all individual participants by measuring the changes in N250 amplitude across single trials. In this step we applied an algorithm that allowed us to trace the time course of learning the target face over trials. We employed the univariate single-trial analysis (Rousselet & Pernet, 2011) and characterized the individual relationships between the N250 amplitude, the number of face presentations, and the performance of the participant.

## 2   METHODS

### 2.1 Participants

Twenty participants were recruited among Polish students of Jagiellonian University, Krakow. All participants were healthy, right-handed Caucasians and had normal or corrected-to-normal vision. They signed informed consent and received a reward for participation. The study was approved by the Jagiellonian University Ethics Committee. Data of four participants were rejected; two, because they finished only part of the experiment, one because of very low response accuracy (< 60%), and one because of low signal-to-noise ratio of the ERP signals.  The final sample consisted of 16 participants (12 females; mean age = 21.5 years; range: 19 to 23).

**2.2 Stimuli**

Two sets of 11 color portraits each of male or female young adult European individuals

were used as stimuli. Pictures were taken from the FACES Lifespan Database for Facial

Expressions (Ebner, Riediger, & Lindenberger, 2010). One of the faces in each set was randomly

selected as target ("Joe" or "Jane") for all participants of the corresponding gender. A picture of the participant (the Own face) was taken right before the experiment on a background similar to the one in FACES pictures. It was manually cropped and normalized to the same dimensions as the other pictures (335 x 419 pixels), using GIMP software. The subset of pictures used for a given participant was gender matched.

**2.3 Procedure**

After signing written informed consent and reading the task instructions, participants were seated in a dimly lit, quiet and electrically-shielded room. A computer screen was placed at a viewing distance of 80 cm. Participants were instructed to sit still, keep quiet, and maintain central eye fixation during the trials.

The experimental procedure was a replication of the Joe/no Joe (or Jane/no Jane) task of

Tanaka et al. (2006). Participants were instructed to monitor the centrally presented faces, where the center of the nose was always in the center of the screen, and indicate whether the face was the target or not by pressing a right or left button using the index and middle finger of the right hand.

At first, participants were familiarized with the Joe or Jane target face. Although this part is crucial for the process of learning, the original description by Tanaka et al. (2006) is a bit vague: "Subjects were then shown the target Joe (Jane) face, and asked to study it." (P. 1490). Thus, we decided to present the Joe (Jane) face on the screen right before the experiment, for 10 to 60 seconds depending on the participant's needs. Although it limits the possibility to analyze the initial phase of face learning, we decided to stick with the (presumably) original procedure. One difference from the original procedure is that we used the same Joe/Jane face for all male/female participants to avoid additional variability connected with physical face characteristics.

The experiment was implemented with E-Prime software. Faces covered around 3.0° x 4.8° visual angle. Each trial began with a 500-ms presentation of a white fixation cross. Then, a face from the set matching the participant's gender was presented for 500 ms, after which the screen went blank for 500 ms; finally, the question "Joe?" ("Jane?" for female participants) was displayed for 500 ms. Participants were instructed to respond only after the question was displayed on the screen. Trials with responses preceding the question were excluded from analysis; if more than one response was given, only the first one was considered. No feedback about performance was given.

Each of the 36 experimental blocks (one more than in Tanaka et al., 2006) consisted of 24 trials (2 presentations each of 10 Other faces, 2 presentations of the Joe/Jane face, and 2 presentations of the participant's Own face). The order of stimuli was pseudo-randomized and different in each block and participant. Joe/Jane and Own faces never repeated immediately to avoid repetition effects (Tacikowski et al., 2011). This presentation regime implies that the intervals (intervening Own and Other faces)  between Joe/Jane faces during the experiment varied across participants.  In total, there were 864 trials per participant (72 Joe trials, 72 Own trials, plus 72 times 10 Other trials) with self-paced breaks between blocks. In total, the task took about 30 min.

## 2.4 EEG/ERP methods

The EEG signal was recorded using the BrainProducts ActiChamp amplifier with 2500 Hz sampling frequency. There were no additional filters applied during recording (the device itself has bandwidth DC – 7500 Hz). The BrainProducts ActiCap was used with 64 active electrodes located according to the 10-10 standard (Acharya, Hani,  Cheek, Thirumala, & Tsuchida, 2016) but oversampling the posterior scalp regions. The Cz electrode was used as initial common reference. The EOG was recorded from a passive electrode located at the canthus of the left eye. All impedances were kept below 10 kOhm. As suggested by the amplifier manufacturer, active noise cancelling was deactivated because active electrodes were used in an electrically shielded room. A Chronos multifunctional response and stimulus device (Psychology Software Tools, USA) was used to collect responses from participants and correct stimulus display times according to the markers from a photodiode mounted on the screen (differences up to 15 ms).

Offline, corrupted channels (3 channels at most) were replaced by spherical spline interpolation (Srinivasan, Nunez, Tucker, Silberstein, & Cadusch, 1996). The EEG signals were digitally low-pass filtered at 40 Hz, using 3rd-order zero-phase forward-backward digital Butterworth filtering (Gustafsson, 1996) with a slope of 6 dB, re-calculated to a common average reference (CAR), and down-sampled to 250 Hz. As the electrodes in our setup were not evenly distributed on the scalp (20 in the anterior and 44 in the posterior scalp), the common average for all electrodes was calculated on a subset of 31 evenly distributed electrodes (Fp1, Fp2, F7, F3,

Fz, F4, F8, FT9, FC5, FC1, FC2, FC6, FT10, T7, C3, C4, T8, CP5, CP1, CP2, CP6, TP9, P7, P3, Pz, P4, P8, TP10, O1, Oz, O2).

The EOG was filtered using ICA decomposition by finding and removing components that were highly correlated with the EOG signal (Hyvärinen & Oja, 2000). Epochs were extracted for time windows between 100 ms before and 900 ms after stimulus onset and corrected with respect to 100-ms pre-stimulus baselines. Trials with activity ranges > 100 μV within any channel or with incorrect responses were discarded from further analyses. All operations were performed using Python 3 and the MNE package (Gramfort et al., 2013).

## 3   RESULTS

### 3.1 Behavioral results

The average accuracy of responses for the 18 participants that completed the whole experiment (including two participants rejected for ERP analyses alone) was high ($M = 91.45\%$, $SD = 11.47\%$) with significantly higher values in the second half of the experiment ($M = 95.66\%$, $SD = 5.76\%$) than in the first half ($M = 87.25\%$, $SD = 18.99\%$) according to the Wilcoxon signed-ranked test ($Z = 3.03$, $p = .002$). The non-parametric test was used because the normal distribution, required for repeated measures ANOVA, was violated according to the Shapiro-Wilk test ($p = .034$). As compared to the first half of the experiment, in the second half, most participants (except for three) achieved greater or equal accuracy for the Joe/Jane faces (Table 1). Average accuracies for both, Joe/Jane (95.11%), as well as Own (93.98%) faces were significantly higher than for Other faces (90.84%; vs. Joe/Jane: $Z = 2.72$, $p = .007$; vs. Own: $Z = 3.29$, $p < .001$). Mean reaction times (for correct trials only) over all conditions were significantly shorter in the second than first half ($M = 1142$ vs. 1186 ms, $SD = 102$ vs. 78 ms) of the experiment ($Z = 3.07$, $p = .002$). These behavioral results confirm the findings of Tanaka et al. (2006) and show that participants effectively learned during the experiment, as a group but also in most individual cases.

**TABLE 1** Accuracy of responses to Joe/Jane faces (in %) for 18 of 20 participants.

|  | 1 | 2 | 3 | 4 | 5 | 6 | 7 | 9 | 11 | 12 | 13 | 14 | 15 | 16 | 17 | 18 | 19 | 20 |
|---|---|---|---|---|---|---|---|---|---|---|---|---|---|---|---|---|---|---|
| 1st | 95 | 100 | 100 | 100 | 100 | 94 | 94 | 100 | 100 | 92 | 97 | 81 | 97 | 36 | 97 | 97 | 89 | 90 |
| 2nd | 97 | 97 | 94 | 100 | 100 | 100 | 97 | 100 | 100 | 100 | 100 | 100 | 100 | 94 | 100 | 100 | 100 | 85 |
| avg | 96 | 99 | 97 | 100 | 100 | 97 | 95 | 100 | 100 | 96 | 99 | 90 | 99 | 65 | 99 | 99 | 94 | 88 |

$1^{st}$ and $2^{nd}$: first and second half of experiment; avg.: all trials.

## 3.2 ERP results

### 3.2.1 Replicating Tanaka et al. (2006)

Following Tanaka et al. (2006), two regions of interest with 6 electrodes each were selected for measuring N250 amplitude in the right and left hemispheres (TP10, P8, P10, PO8, PO10, O2 and TP9,P7, P9, PO7, PO9, O1). ERP waveforms were obtained by averaging over all electrodes within each ROI, per condition and each half of the experiment (Fig. 1). The resulting ERP waveforms resembled the ones shown in (Fig. 2 of Tanaka et al., 2006), especially for Other faces, but they also show some differences. For present purposes, the most important difference is that the N250 "dip" was visible only for ERPs to Joe/Jane faces but not at all for ERPs to Own faces. Furthermore, this dip was present in both halves of the experiment and not just in the second half as in Tanaka et al. (2006). In addition, there was a negative-going shift in the ERPs between 400 and 600 ms post-stimulus that was not present in the waveforms shown by Tanaka et al. (2006).

As in Tanaka et al. (2006) the average amplitude of the N250 was measured in the averaged ERPs in each ROI between 230 and 320 ms and submitted to a repeated measures analysis of variance (ANOVA) with factors Condition (Own, Joe, Other faces), Experiment Half

(First, Second), and Hemisphere. Degrees of freedom were adjusted for sphericity violations according to the conservative Greenhouse–Geisser procedure (Winer, Brown, & Michels, 1991).

Replicating Tanaka's results, there were clear N250 amplitude differences between conditions, $F(2,15) = 14.06$, $MSE = 7.84$, $p < .001$. However, confirming visual impressions,

condition effects were not qualified by an interaction with experiment half ($F(2,15) = 0.91$, $MSE$

$= 1.89$, $p = .38$). More specifically, the null hypothesis of equal N250 amplitudes cannot be

rejected at the alpha $= 0.01$ level according to Bonferroni corrected follow-up contrasts. The strip plots and distributions of the differences in Figure 3 show the detailed interaction of condition and part. The contrasts revealed also that N250 to Joe faces significantly differed from Own faces in both parts of the experiment, whereas in Tanaka et al. (2006) amplitudes had been indistinguishable in the second half. Furthermore, we could not replicate the left-hemispheric asymmetry reported by Tanaka et al. (2006), $F(1,15) = 2.23$, $MSE = 8.67$, $p = .16$.

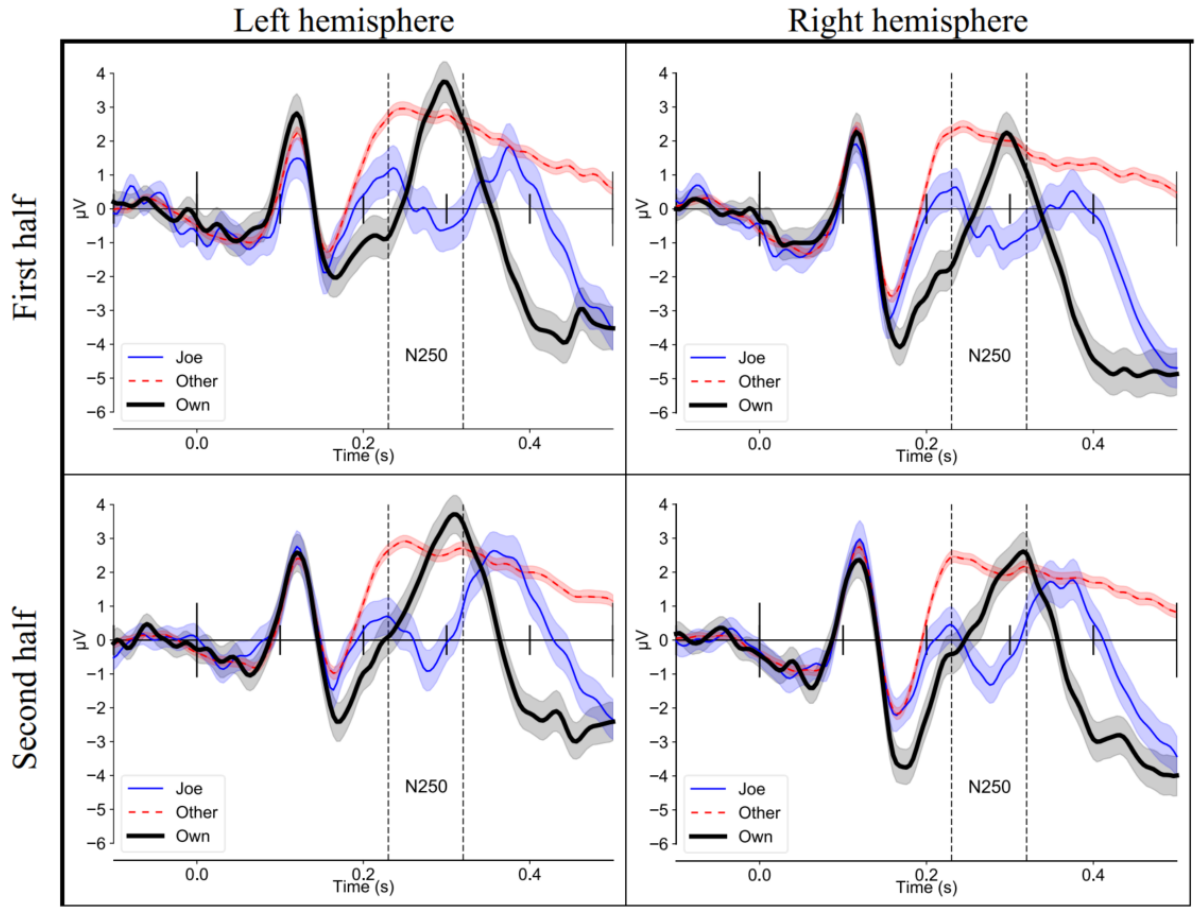

**FIGURE 1** Grand-average ERPs for the two halves of the experiment averaged across channels at left and right regions of interest (the equivalent of Fig. 2 of Tanaka et al., 2006).

### 3.2.2 Exploring an alternative trial division – 1/3 to 2/3

The non-replication of larger N250 amplitudes in the second experimental half as reported by Tanaka et al. (2006) is a serious concern for the main aim of the present study, that is, analyzing the fine-grained time course of N250 across learning a novel face. Possible reasons for the non-replication despite our efforts to adhere to the original procedure, may be differences in the participant samples, the choice of the particular Joe/Jane faces, or the duration of the presentation of these faces at the beginning of the experiment. All these factors might have facilitated the learning and confined it to the first few trials of the first experiment half.

Therefore, we searched for an alternative split by checking the difference in means between experimental parts in steps of 1/24 of trials (i.e., 1/24 vs 23/24, 2/24 vs 22/24, etc.). Only the divisions providing a reasonable number of trials (at least 20 per participant (Luck, 2006)) for grand ERP averages to be meaningful were considered. In the second step, we chose from the candidates the split giving the maximum difference of means. It turned out that split 1/3-2/3 was the best over all the splits explored. Grand averages of this new division are shown in Figure 2.

ANOVA of N250 amplitude with the new division (factor Experiment part (1st 1/3 vs. last 2/3 of trials) revealed a significant interaction of condition and experiment part, $F(2,15) = 5.73$, $MSE = 1.56$, $\underline{p} = .014$.

Follow-up contrasts (Bonferroni corrected for multiple comparisons, $p < .01$) revealed that N250 amplitudes for targets (Joe or Jane faces) were significantly more negative in the last 2/3 than in the first 1/3 of trials ($M = -0.51$ vs. $0.28$ μV) of the experiment (Fig. 3). The strip plots and distributions of the differences in Figure 3 show the detailed interaction of condition and part.

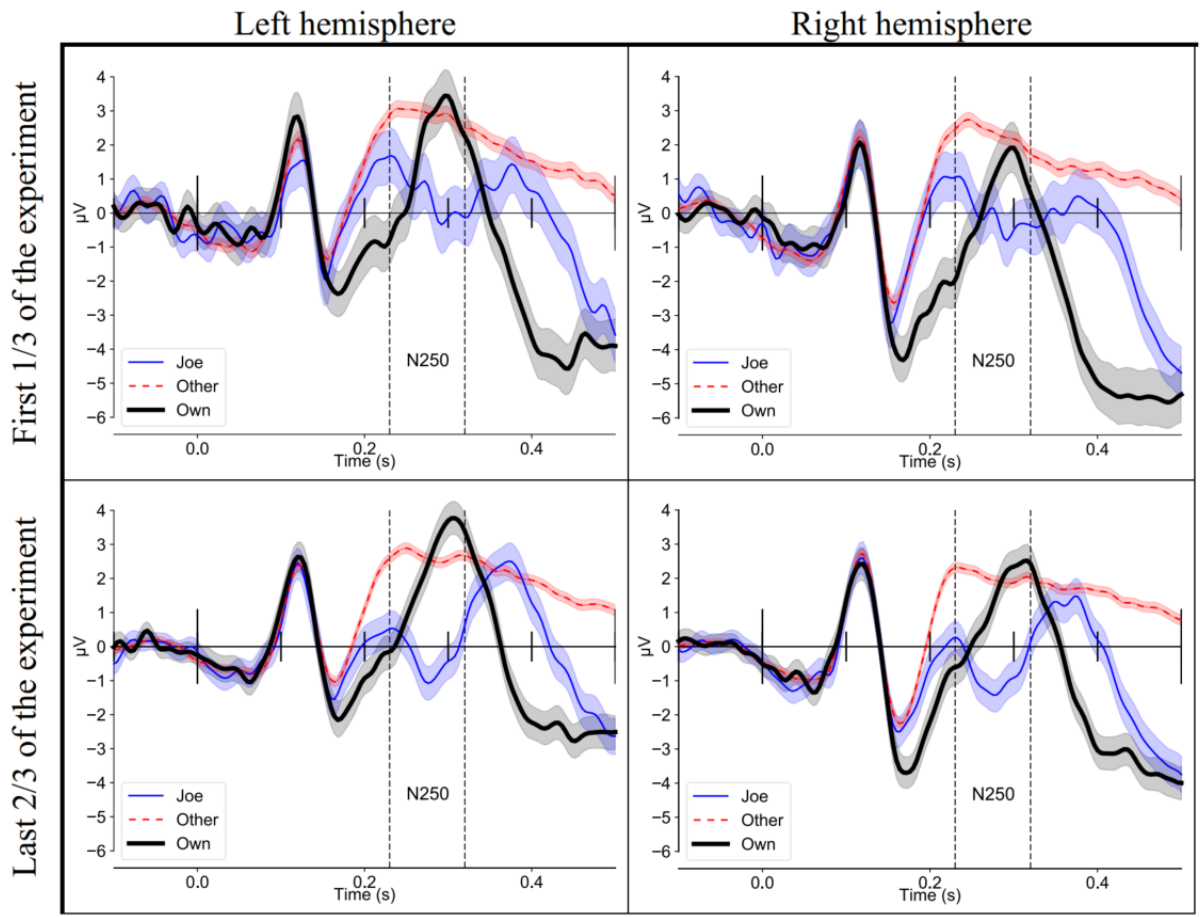

**FIGURE 2** Grand-average ERPs for the first 1/3 of trials (top) and last 2/3 of trials (bottom) of the experiment for the left and right regions of interest of the N250 component (marked with vertical lines).

Using the alternative division allowed to replicate the crucial effect of more negative N250 amplitudes to Joe/Jane faces in the later part of the learning process. These results indicate that similar learning effects were present as in the study of Tanaka et al. (2006) but shifted towards earlier parts of the experiment and enabled the analysis of the time course of face learning based on N250 amplitude in single trials, to be reported next.

It is worth mentioning that the same analysis for N170 component in the time range of 130-200 ms did not reveal any significant interaction of condition and experiment part, supporting the notion of N250 as specific familiarity effect, and confirming the findings from the referenced articles.

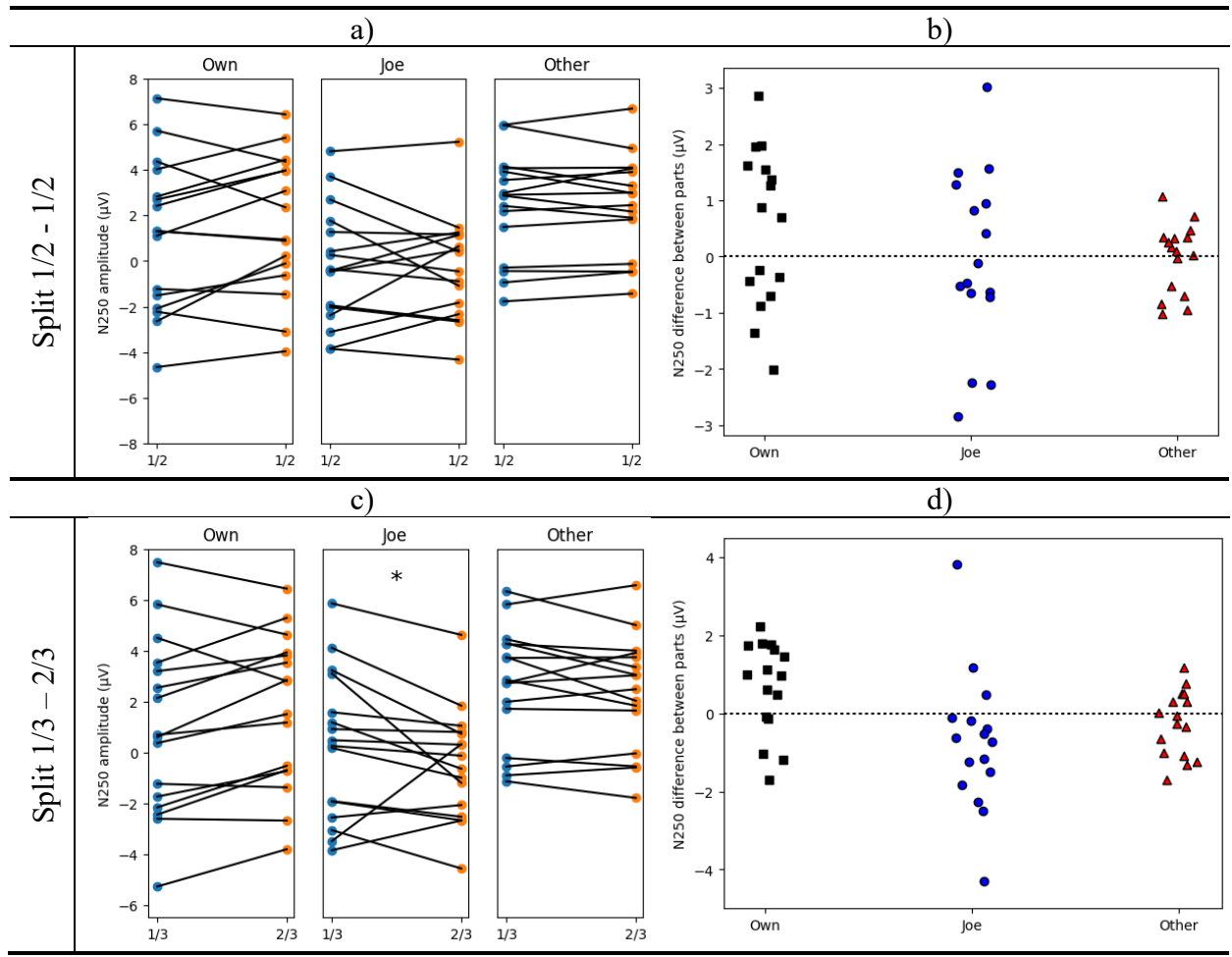

**FIGURE 3** Mean N250 amplitudes across all electrodes of interest for the Own, Joe(/Jane) and Other conditions and their changes for different splits of the experimental trials. Panels a), c) show strip charts of linked observations for each participant for different splits. Panels b), d) show a distribution of N250 amplitude differences between parts. The asterisk indicates a reliable difference between the two parts of the experiment for the Joe condition.

## 3.3  Exploring single trial analysis of N250 amplitude across time

Our main research goal was to quantitatively characterize the changes of N250 amplitude during the course of learning a face, using single-trial ERP analysis (Blankertz, Lemm, Treder, Haufe, & Müller, 2011). We also wanted to be able to describe the learning process on the levale of individual participants. Besides the described preprocessing, the mean N250 amplitude for each epoch and all electrodes of the N250 ROI was calculated in the time range 230 to 320 ms post-stimulus. A sample time series of the N250 amplitude to target faces is shown in Figure 4.

In a first approach, we considered the trends across face learning, as reflected in the time course of N250 amplitude, using different regression models like linear, quadratic, exponential, GLM, GLMM and GAM for each participant. Unfortunately, the adjusted R-squared coefficients of determination for all models fits were very low, reaching at most 15 percent, making it problematic to find a single quantitative function to model the time course of face learning. Therefore, we took another approach and specified our goal into detecting a significant change (change point or break-point) in the linear trend across N250 amplitudes. The time point of course change in N250 amplitude might signal the transition from an early phase of acquiring face representations to a later phase where these representations are already more stable.

First, we applied segmented regression from the *segmented* R package (Muggeo, 2003). It is an iterative procedure in which a standard GLM is estimated and then a broken-line relationship is added by re-fitting the overall model. The iterations start from an arbitrary starting point which isn't straightforward to select and strongly affects the result when dealing with noisy data (we use the default setting – median of the trial numbers). Also, it assumes that straight lines for each segment must intersect exactly in the change point. Additionally, in the determination of the statistical significance of potential break-points, the requirement of data in parametric tests was not fulfilled in our data set precluding the application of this method. Despite this, we reported the results from the *segmented* package in Table 2, in order to compare them with the results of our new proposed method.

To achieve our goal we propose the new two-step explorative-confirmative non-parametric method (not requiring normality) based on the permutation tests. In the first explorative step, all points are searched consecutively in order to select a candidate, which is the point for which the estimated Cohen's effect value is maximum. Then, in the second confirmative step, permutation test checking the statistical significance of the candidate change point is performed. This permutation test is based on the proposed test statistic which enables to detect trend changes in such noisy time series. The detailed description of these steps follows.

We aimed to find a point in time during face learning where a first negative-going linear trend towards larger N250 amplitudes was replaced by a trend with a less negative slope, that is, we searched for a "change point" between two linear trends during face learning.

Let us denote $n$ paired experimental data observations by $(x_i, y_i)$ for $i = 1, 2, \ldots, n$. We will consider a simple linear regression model

$$y = \beta_0 + \beta_1 x + \varepsilon \tag{1}$$

where $y$ is the dependent variable (here, the N250 amplitude), $x$ is the independent or explanatory variable (here the trial number taking values from the set {1,2,..., 864}), $\beta_0$ and $\beta_1$ are parameters of the linear model and $\varepsilon$ is an error term.

Cohen (1988) defines an effect size $d$ as follows

$$d = \frac{m_2 - m_1}{\sigma}$$

where $m_1$ and $m_2$ are populations means under considerations expressed in raw (original) measurement units and $\sigma$ is the standard deviation of either population of measurements (under $H_0$ the variances are equal). The parameter $\sigma$ is estimated by the following formula

$$\hat{\sigma} = \sqrt{\frac{(n_1-1)\,S_1^2 + (n_2-1)\,S_2^2}{n_1 + n_2 - 2}}$$

where $S_j^2 = \frac{1}{(n_j-1)} \sum_{i=1}^{n_j}(x_{ji} - \bar{x}_i)^2$ for $j = 1, 2$.

Let us consider two sets $\mathbf{S}_1 = \{(x_i, y_i): i = 1, 2, \ldots, k\}$ and $\mathbf{S}_2 = \{(x_i, y_i): i = k+1, k+2, \ldots, n\}$ where $k = 10, 11, \ldots, 50$. Based on these two sets we get two regression lines with slopes $\beta_{11}$ and $\beta_{12}$. Let $d(k)$ now measures the estimated (with permutation method) Cohen's effect of the population slope of the linear regression. The change point $k^*$ is selected separately for each participant using the formula:

$$k^* = \min_{k \in K} \left\{ k : d(k) = \max_{k \in K} d(k) \right\}$$

where $K = \{10, 11, \ldots, 50\}$.

The significance of the change point is tested using the permutation test. We consider the null hypothesis that the dependence between $x$ and $y$ can be described by a linear model (1). The alternative hypothesis, that there are two linear models for sets $\mathbf{S}_1 = \{(x_i, y_i): i = 1, 2, \ldots, k^*\}$ and $\mathbf{S}_2 = \{(x_i, y_i): i = k^*+1, k^*+2, \ldots, n\}$ with different slopes $\beta_{11}$ and $\beta_{12}$, is considered (where $k^*$ is the experimental point, which maximizes Cohen's effect, $n$ is the number of all participant's trials with Joe/Jane face). Formally, the null hypothesis and the alternative can be written as follows:

$H_0$: The dependency between $x$ and $y$ can be written as a linear model (1)

$H_1$: There is a change in trend at point $k^*$. Two linear models are considered

$y = \beta_{01} + \beta_{11}x + \varepsilon$ for $i = 1, 2, \ldots, k^*$ (2)

$y = \beta_{02} + \beta_{12}x + \varepsilon$ for $i = k^*+1, k^*+2, \ldots, n$ (3)

where $\beta_{12} > \beta_{11}$. The idea of a trend change is presented in Figure 4.

Let us assume the significance level $\alpha = 0.05$. The test statistic should take different values under the null hypothesis and for the alternatives we care about. We consider the following test statistic

$T = b_{12} - b_{11}$ (4)

where $b_{11}$ is the slope of the linear regression obtained for $i = 1, 2, \ldots, k^*$ and $b_{12}$ is the slope of the linear regression obtained for $i = k^*+1, k^*+2, \ldots, n$ ($b_{11}, b_{12}$ are the estimators of the corresponding parameters of the linear regression model). For the original data (i.e. without permutation, see below) the value $T_0$ of the statistic is calculated (according to formula (4)). Under $H_0$ on the basis of the regression model (1) the residuals are calculated. Then the residuals are permuted $N = 10000$ times and for each case two models of regression are obtained and the value of the test statistic is calculated according to formula (4) (resulting in $N$ values $T_1, T_2, \ldots, T_N$ of the $T$ statistic. The empirical distribution of the test statistic $T$ under $H_0$ is obtained. If the estimated $p$-value, given by Good (2006)

$$p.value \approx \frac{card\{i: T_i \geq T_0\}}{N}$$

is less than the significance level $\alpha$, the null hypothesis is rejected.

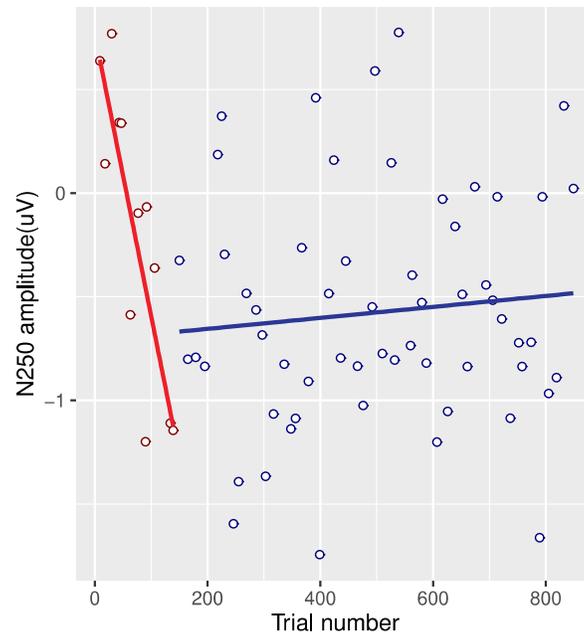

**FIGURE 4** Hypothetical change point between two linear regression models (data of Participant 6). The red line shows the best-fitting linear regression for points 1 to k*, whereas the blue line shows the best-fitting linear regression for points k*+1 to n.

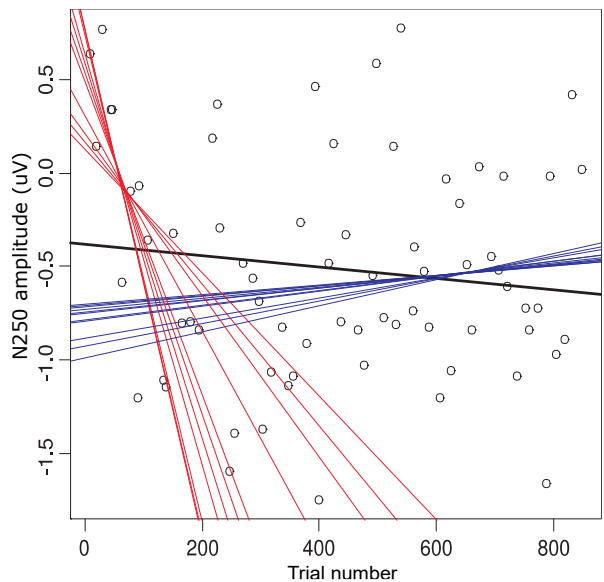

**FIGURE 5** Linear regressions for N250 amplitudes of participant 6. Black line: linear regression for all observations. Red and blue lines: Linear regressions for observation 1 to k and k+1 to 64, respectively, where k = 10, 11, …, 20.

The procedure described above was applied to the single trial N250 amplitudes to Joe/Jane faces in 16 participants with valid ERP data, to search for a change point in linear trends. The permutation test for equal slopes in two linear models for given sets $S_1$ and $S_2$, as described above was used, looking for the change point for $k^*$, obtained for given participant. In 13 data sets a change point in linear models was detected. Specifically designed test statistics maximizing Cohen' effect size resulted in unequivocal identification of change points in trends. This, in turn, could be a justification of bi-phasic division of time course of face learning.

In most cases the change point was detected at the beginning of the trial series ($k^*$ between 10 and 21), in 4 cases in the middle phase ($k^*$ between 28 and 39) and in 3 cases at the end ($k^*$ between 45 and 50). The change points $k^*$ with corresponding $p$-values and Cohen's $d$ for each analysed participant are presented in Table 2. Figure 6 presents the trends for all participants, ordered as in the Table 2.

Comparison presented in Table 2 shows that all the change points detected by our method ($k^*$) are close to the change points detected by the segmented method, and they all fall into 95% confidence intervals calculated by the segmented method..

**TABLE 2** Change points detected by our method (k*) and the *segmented* package ($k^{seg}$) for all participants. Together with Cohen's effect d and p-value for our method, and 95% CIs computed by the *segmented* package.

| | our method | | | *segmented* package | |
| --- | --- | --- | --- | --- | --- |
| Participant # | k* | d | p-value | $k^{seg}$ | CI (95%) |
| 17 | 10 | 3.707 | .0034 | 14 | [4, 24] |
| 6 | 12 | 4.706 | .0002 | 10 | [4, 15] |
| 3 | 14 | 3.045 | .0145 | 9 | [-2, 21] |
| 2 | 16 | 4.028 | .0019 | 10 | [3, 18] |
| 15 | 17 | 3.261 | .0103 | 11 | [-4, 27] |
| 9 | 21 | 4.293 | .0015 | 19 | [6, 32] |
| 5 | 28 | 3.883 | .0031 | 25 | [14, 37] |
| 18 | 29 | 3.136 | .0127 | 27 | [8, 45] |
| 7 | 34 | 5.006 | .0003 | 31 | [22, 40] |
| 13 | 39 | 2.674 | .0306 | 48 | [27, 68] |
| 19 | 45 | 3.886 | .0024 | 37 | [28, 45] |
| 11 | 49 | 4.359 | .0008 | 51 | [38, 64] |
| 12 | 50 | 3.513 | .0059 | 55 | [44, 66] |
| 4 | none | 1.656 | .1219 | 63 | [34, 92] |
| 14 | none | 1.196 | .2012 | 40 | [32, 49] |
| 20 | none | 1.924 | .0868 | 28 | [9, 46] |

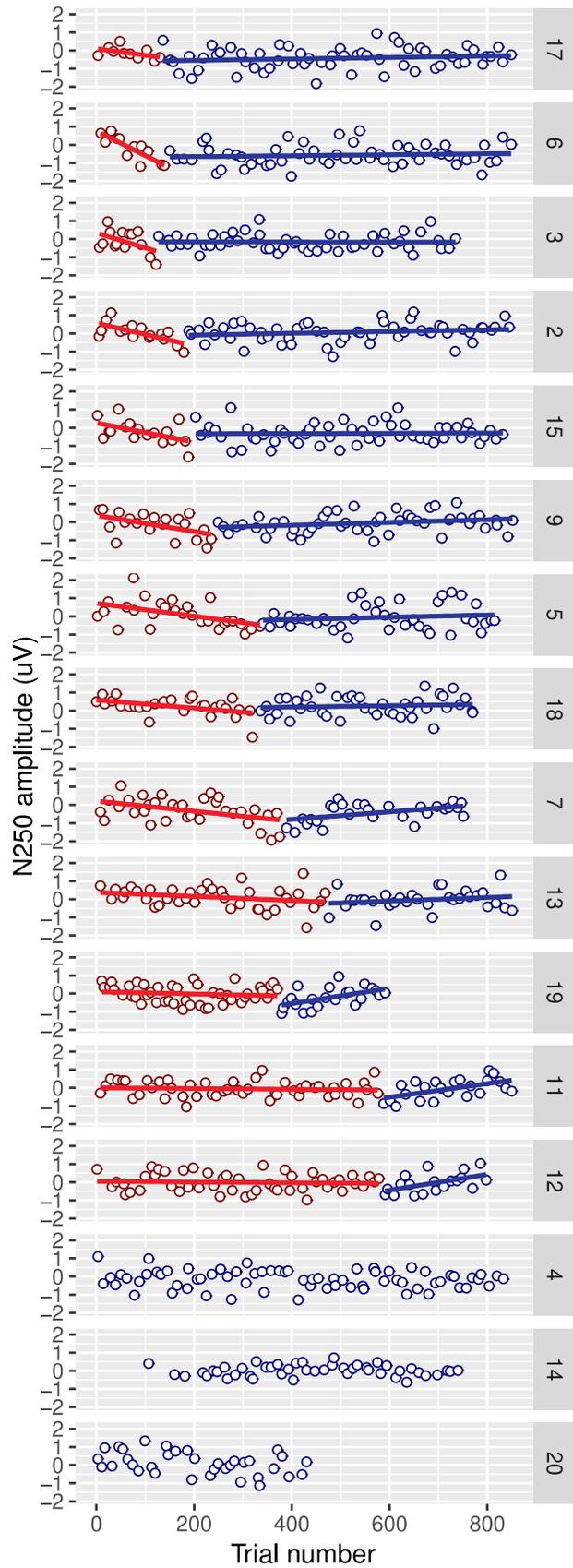

**FIGURE 6** Time series of N250 amplitudes for all participants, sorted according to the timepoint k, where a significant change between two linear trends were detected. The three participants at the bottom did not show any significant linear regressions.

**3.4 Relation of performance and change point**

The relation between the performance (percent correct in detecting Joe/Jane pictures) and the change point was considered. Since performance seemed to be of interest mainly at the beginning of the experiment, only the first half trials was taken into account.

The Pearson correlation coefficient between performance in the first half of the experiment (Table 1) and $k^*$ (Table 2) is equal to -.41. For testing the significance of the Pearson correlation coefficient permutation test for correlation coefficient was used. The null hypothesis had a form

$$H_0: \rho = 0$$

against the alternative

$$H_1: \rho < 0$$

where $\rho$ is the population correlation coefficient.

The $p$-value was equal to .08. This correlation coefficient is significant at the level of $\alpha = 0.1$. It could be said that the greater value of the performance leads to the earlier change point in the learning process.

**4   DISCUSSION**

The present study had two aims, firstly, to replicate the findings of Tanaka et al. (2006) that the N250 increases in amplitude from a first to a second part of a long series of face presentations, leading to increasing familiarity. Second, we explored how to characterize these changes at the level of single trials and individual participants.

**4.1 Replicating Tanaka et al. (2006)**

In our replication of the study of Tanaka et al. (2006), we monitored ERPs while participants viewed faces including their own face (familiar face), an experimenter-specified unfamiliar target face (Joe or Jane), and 11 other unfamiliar nontarget faces. The increasing

N250 amplitudes in averaged ERPs between two parts of the experiment, reported by Tanaka et al. (2006) was observed only after dividing the target trials into the first 1/3 and the last 2/3 of samples, rather than first and second half. This may suggest that on average, our participants learned faster, our Joe and Jane faces were easier to learn, and/or the pre-learning phase was more efficient from that of Tanaka et al. (2006). The last case could be a good starting point for a new study on the impact of pre-learning regimes on face learning characteristics.

Whatever the reasons, why the overall difference between early and late trials emerged earlier in the present study as compared to the report by Tanaka et al. (2006), it does lend support to the idea that N250 amplitude is sensitive to the acquisition of memory representations of a new face image. The N250 amplitude for Other faces was stable across both parts of the experiment. The results for the non-Joe faces indicate that the ERP changes to target faces were no artifact of time-on-task such as fatigue but instead were indeed due to learning.

A more realistic face learning procedure would involve the recognition of the 'Joe/Jane' face across a larges range of image variations of the same person (e.g., Kaufmann et al., 2009); this may be especially relevant as it is surprisingly difficult to recognize unfamiliar faces if different views of the same faces are used – as unavoidable in real life settings (Bruce et al., 1999). The findings from unfamiliar-face-matching described in (Young & Burton, 2017) show that people have a perceptual problem with recognizing that different images can represent the same unfamiliar face identity. Therefore, as our main goal was the detection of significant change in trend during the acquisition of face representations for initially unfamiliar faces, we decided to use identical images of the same persons. However, it would be a promising extension of the present approach to apply our method to unfamiliar face learning across image variability in future research.

Apart from apparently faster learning of the target faces in the present study as compared to Tanaka et al. (2006), we also found that Own faces did not elicit a comparable N250 amplitudes as the target faces in the second part of the experiment. It is not clear how to explain this discrepancy. The Own faces in the present study had been taken immediately prior to the experiment and may have been relatively unfamiliar for the participants; however, it is not described how Tanaka et al. obtained the pictures for their own condition.

## 4.2. Tracing the time course of N250

Our direct replication attempt of Tanaka et al. (2006) failed but after a modest adaption of separating the trials into 1/3 vs. 2/3 rather than ½ vs. ½, we were able to confirm the essential finding of an increasing N250 amplitude with repeated target face repetition, supporting the interpretation of the N250 as a correlate of face familiarity and acquiring perceptual expertise (e.g. Andrews et al., 2018; Kaufmann et al. 2009; Pierce et al. 2011; Schulz et al. 2012; Tacikowski et al., 2011). This conceptual replication provides the basis for the main aim of our study – to develop an algorithm that allows tracing the time course of learning a face over subsequent trials, thus to perform a single trial analysis on N250 responses.

There was very large variance in single trial N250 amplitudes. An attempt to use a single type of univariate linear or exponential regression function across the full time-course of the experimental trials did not give the expected results. The variance explained ($R^2$ coefficients) was very small, in the order of a few percent only. This result suggests that finding a quantitative univariate model of the time course of face learning is very difficult.

As an alternative, we applied a permutation tests-based approach to design a new statistical method to detect the "*change point*" during the time course of face learning. This idea is based on the assumption that face learning may take place in two phases, an initial phase of fast learning and a later phase with slower or no learning. This assumption was encouraged by the initial observation that dividing the 72 target trials of the learning run into a 24 versus 48 was better able to show an N250 amplitude difference than two equal halves of 36 trials each.

The experimental results of 16 participants were available. The number of usable trials per participant was typically between 67 and 71; in three cases it was smaller: for participants 7, 14, and 20, it was 62, 46 and 35 measurements, respectively. In order to detect patterns and trend changes in the face recognition learning process we used *permutation tests*. These tests are characterized by low requirements on the distribution of variables and sample size. Permutation methods are both feasible and practical, have high power and are ideal for small datasets. These tests can also be applied to compare two observation series for a single participant. Permutation tests can be defined for any selected test statistic. Researchers have the option to choose a wide variety of test statistics. Using permutation tests the key issue is to choose the test statistic, which should take different values under $H_0$ and if $H_0$ is false. For the present case the null hypothesis assumed that there is no change in linear trend between a first and a second part of the experiment. The alternative hypothesis was that the process can be described by two linear trends. Directional hypotheses that the slope coefficient in the second phase is greater (more positive) than the (presumably negative) slope coefficient in the first phase of the learning process were used. As test statistic, the difference between the slope coefficients of the trends in the first and second phase of the face recognition learning process was used. The rejection region of this test was right-sided. The permutation test confirmed that in most participants two phases in the face recognition learning process can be distinguished,

We chose the time point with the maximum estimated effect size (Cohen's d) as best estimate for the change point between linear trends. This procedure unambiguously indicated the point of trend change in 13 of the 16 analysed participants. The detected change points in linear trends differed between participants and varied from 12 to 50 presentation of the Joe/Jane face. No statistically significant changes in trend were detected in three participants (No.  4, 14, and 20).  In cases when there is no detectable change in linear trend, there are three options. (1) Learning is linear, which should be reflected in a significant linear trend over the whole experiment. (2) there is no learning, or (3) learning takes place at first encounter. In the three cases without significant change of linear trends, there was also no significant overall (linear) trend. Therefore, we might conclude that in these cases there was little change over the whole experiment. As our method relies on permutations, there may be slight differences in the estimated values for consecutive simulations with different random seeds. However, the statistical significances of the change points were stable across multiple simulations with different random seeds (e.g. the change points for participants 4, 14, 20 were never statistically significant). Moreover, the experimental comparison with the segmented  method, shows that all the change points detected by our method fall into into 95% CIs calculated by the segmented method. This can be an auxiliary argument for a good generalizability of our method.

Our single-trial analysis is based on the assumption that face learning can be conceived to usually consist in two linear trends. The assumption of a *linear change* that stops after a particular number of target presentations is likely to be only a crude approximation for more non-linear developments. Nevertheless, this approximation constitutes a progress over the previous approaches of averaging ERPs over blocks of learning. The present approach allows to identify change points that differ between individuals.  Although the relationship of the individual change point of the trend in N250 amplitude with performance should be considered as tentative, it indicates that it will be fruitful to attempt more systematic studies, ideally with independent tests of face learning performance and change points of the N250 amplitude trends, complementing previous studies on the neural correlates of individual differences in face cognition (Herzmann, Kunina, Sommer, & Wilhelm, 2010; Kaltwasser, Hildebrandt, Recio, Wilhelm, & Sommer, 2014; Nowparast Rostami, Sommer, Zhou, Wilhelm, & Hildebrandt, 2017).

In our study we took the seminal study of Tanaka et al. (2006) as a reference and starting point, which offers the safety of a replication of the most important finding, that is, the increase of N250 amplitude across two blocks of presentations. Using these findings as a stepping stone, we moved on to a more fine-grained analysis, be combining single trial ERP analysis with permutation testing of different linear trends. This approach goes beyond the study of the averaged brain response by analyzing all single trial responses for each participant separately. It allows to interpret individual differences by quantifying effects within and between participants, providing data descriptions at a level needed to build efficient models of face learning (Pernet, Sajda, & Rousselet, 2011). The new method allows to study the time course of the acquisition of representations for novel faces and enlarges the toolbox for investigation neural correlate of face (or object) learning and may be useful for applied contexts.

**ACKNOWLEDGEMENTS**

This research was supported by statutory funds of Department of Applied Informatics, Silesian

University of Technology, Gliwice, Poland (grant SUBB/RAu7/2020).

We declare no conflict of interest.